\documentclass[pra,twocolumn,showpacs,superscriptaddress]{revtex4-1}
\usepackage{amsmath}
\usepackage{amssymb}
\usepackage{graphicx}
\usepackage{color}

\begin{document}

\title{Generalized Sub-Schawlow-Townes Laser Linewidths Via Material Dispersion}

\author{Jason Cornelius Pillay}

\author{Yuki Natsume}

\affiliation{Division of Physics and Applied Physics, School of Physical and Mathematical Sciences, Nanyang Technological University, Singapore 637371, Singapore}

\author{A.~Douglas~Stone}

\affiliation{Department of Applied Physics, Yale University, New
  Haven, Connecticut 06520}

\author{Y.~D.~Chong}
\email{yidong@ntu.edu.sg}

\affiliation{Division of Physics and Applied Physics, School of Physical and Mathematical Sciences, Nanyang Technological University, Singapore 637371, Singapore}

\affiliation{Centre for Disruptive Photonic Technologies, Singapore 637371, Singapore}

\begin{abstract}
A recent $S$ matrix-based theory of the quantum-limited linewidth,
which is applicable to general lasers, including spatially non-uniform
laser cavities operating above threshold, is analyzed in various
limits. For broadband gain, a simple interpretation of the Petermann
and bad-cavity factors is presented in terms of geometric relations
between the zeros and poles of the $S$ matrix. When there is
substantial dispersion, on the frequency scale of the cavity lifetime,
the theory yields a generalization of the bad-cavity factor, which was
previously derived for spatially uniform one-dimensional lasers.  This
effect can lead to sub-Schawlow-Townes linewidths in lasers with very
narrow gain widths.  We derive a formula for the linewidth in terms of
the lasing mode functions, which has accuracy comparable to the
previous formula involving the residue of the lasing pole.  These
results for the quantum-limited linewidth are valid even in the regime
of strong line-pulling and spatial hole-burning, where the linewidth
cannot be factorized into independent Petermann and bad-cavity
factors.
\end{abstract}

\maketitle

\section{Introduction}

One of the oldest problems in laser physics is the characterization of
the quantum-limited laser linewidth.  In their seminal paper on the
theory of the laser, Schawlow and Townes derived the formula \cite{ST}
\begin{equation}
  \delta \omega_{\mathrm{ST}} = \frac{\hbar\omega_0\gamma_c^2}{2P},
  \label{omegast}
\end{equation}
where $\omega_0$ is the frequency of the laser mode, $\gamma_c$ is the
linewidth (FWHM) of the corresponding passive cavity resonance, and
$P$ is the output power.  Several corrections to this result were
found by subsequent researchers: (i) an excess noise factor arising
from incomplete population inversion in the gain medium; (ii) the
Petermann factor, which describes excess noise due to mode
non-orthogonality
\cite{Petermann,Haus,Henry86,Siegman,HamelWoerdman,HW2}; (iii) the
Henry $\alpha$ factor, which describes indirect phase fluctuations
from instantaneous intensity changes caused by spontaneous emission
\cite{Henry82}; and (iv) a ``bad-cavity'' factor which reduces the
linewidth when the cavity decay rate is on the order of the gain width
\cite{Lax,Haken,Kolobov,Woerdman,Kuppens1,Kuppens2}.  The first three
factors all broaden the linewidth relative to the basic
Schawlow-Townes result, Eq.~(\ref{omegast}).  The bad-cavity factor,
however, \textit{reduces} the linewidth.  Its origin was originally
attributed to the slowdown of phase diffusion caused by atomic memory
\cite{Lax,Haken,Kolobov}; subsequently, Kuppens \textit{et al.}~gave
an alternative interpretation based on the increase in the laser
cavity's group refractive index due to the frequency dispersion of the
gain medium \cite{Kuppens1}.  This factor deviates significantly from
unity in bad-cavity lasers, whose cavity decay rates are on the order
of the gain width (or polarization dephasing rate), and has been
demonstrated experimentally in a HeNe gas laser
\cite{Kuppens1,Kuppens2}.  It has also recently been re-derived in the
context of quantum cascade lasers, where it yields a small but
measurable correction to the linewidth \cite{liutao}.  Recently, there
have been theoretical proposals to achieve ultra-low linewidth lasers
by exploiting this effect with superradiant gain media
\cite{superrad0,superrad1,superrad2,superrad3,superrad4}.

Recently, two of the present authors developed a theory of the
quantum-limited laser linewidth \cite{linewidth_prl}, based on the
properties of the scattering matrix ($S$ matrix) derived from
Steady-state Ab-initio Laser Theory (SALT)
\cite{salt1,salt2,salt3,spasalt}.  According to this theory, the
cavity decay rate $\gamma_c$ in Eq.~(\ref{omegast}) is replaced by a
generalized decay rate
\begin{equation}
  \gamma_L = \left|\mathrm{Res}(s)\,
  \frac{\Psi_L^\dagger\Psi_L}{\Psi_L^T\Psi_L}\right|,
  \label{generalized decay rate}
\end{equation}
where $\mathrm{Res}(s)$ denotes the residue of the $S$ matrix
eigenvalue, $s$, which diverges at the laser frequency, and $\Psi_L$
is the corresponding $S$ matrix eigenvector.  Note that $\Psi_L$ is
not the lasing mode function, i.e.~it is not the electric field as
function of position, but rather an $N$-component complex vector,
where $N$ is the number of asymptotic scattering channels coupled to
the laser cavity. The quantity denoted $\Psi_L^\dagger\Psi_L$ is the
usual hermitian norm of this vector, which will be set to unity by
convention.  The quantity $\Psi_L^T\Psi_L = \sum_{i=1}^N \Psi_{L,i}^2$
is the biorthogonal norm of $\Psi_L$, a complex number with modulus
less than or equal to unity.  It does not represent the Petermann
factor, $K$, despite its apparent similarity to familiar integral
formulas for the same; in fact, for one-port ($N = 1$) lasers,
$|\Psi_L^T\Psi_L|=1$ even though $K<1$.

SALT describes single- or multi-mode lasing above threshold for
arbitrary laser cavities, and the $S$ matrix used in
Eq.~(\ref{generalized decay rate}) is a non-linear $S$ matrix computed
from the SALT equations as described in \cite{linewidth_prl}.  It
takes into account both the gain competition, spatial hole-burning and
self-saturation effects present above threshold.  It was shown in
Ref.~\onlinecite{linewidth_prl} that Eq.~(\ref{generalized decay
  rate}) incorporates the incomplete inversion and Petermann factors,
and due to the generality of the $S$ matrix approach, it can be
applied to complex modern laser cavity geometries, such as microdisk,
photonic crystal, and random lasers.  By contrast, previous
derivations of the Petermann factor have been specific to
one-dimensional (1D) cavities
\cite{Petermann,Haus,Henry86,Siegman,HamelWoerdman,HW2}, with the
notable exception of a paper by Schomerus \cite{schomerus} which will
be discussed below.

In this paper, we analyze the $S$-matrix linewidth formula further,
and derive additional results which flow from it.  We show that
Eq.~(\ref{generalized decay rate}) also exhibits the bad-cavity
linewidth reduction effect mentioned above
\cite{Lax,Haken,Kolobov,Woerdman,Kuppens1,Kuppens2}.  Before proving
this, we analyze a one-port laser (e.g. a Fabry-P\'erot single-mode
laser), for which the $S$-matrix can be completely described in terms
of the poles and zeros positions. This leads, in Sections
\ref{Petermann section} and \ref{bad cavity section}, to a simple
geometric interpretation of the Petermann and bad-cavity factors in
terms of the motion of the poles and zeros of the $S$-matrix in
response to the pumping of the gain medium. For more general lasers,
including non-1D and/or spatially non-uniform lasers, the full
$S$-matrix theory allows for a more rigorous calculation of the
linewidth.  In Section \ref{Wavefunction Formulation}, we use the
$S$-matrix theory to derive an alternative formula for $\gamma_L$ in
terms of the lasing wavefunction, and show that the bad-cavity
linewidth reduction factor is automatically incorporated.  By
contrast, the bad-cavity factor was derived in
Refs.~\onlinecite{Lax,Haken,Kolobov} using Langevin equations, in
which the spatial variation of the lasing mode is neglected, and in
Ref.~\onlinecite{Woerdman} using a Green's function method specific to
1D cavities.  For Fabry-P\'erot cavities, we show analytically and
numerically that our theory reduces to the earlier results.  In
Section \ref{numerical section}, we present numerical analyses of more
complex lasers, including spatially non-uniform 1D cavities and 2D
cavities.  When spatial hole-burning and line-pulling (due to the
frequency dispersion of the gain medium) are negligible, the
$S$-matrix theory is in good agreement with previous, more approximate
theories, where the bad-cavity factor and Petermann factor are treated
as independent quantities.  In the presence of strong line-pulling or
spatial hole-burning, we find that this factorization breaks down.
The deviations in the linewidth predicted by the $S$-matrix theory,
under these more general conditions, can be tested in future
experimental work.

\section{Geometric view of the Petermann factor}
\label{Petermann section}

Possibly the most-studied correction to the Schawlow-Townes linewidth
formula is the Petermann factor, which accounts for the fact that the
modes of any open system, including laser cavities, are
non-orthogonal.  When spontaneous emission noise is decomposed into
these non-orthogonal modes, there is an excess in the overall noise
level associated with noise correlation in different modes
\cite{Haus}.  This effect was originally discovered and discussed in
the context of transverse modes of gain-guided lasers
\cite{Petermann,Haus,Siegman}, and subsequently extended to
longitudinal modes by Hamel and Woerdman
\cite{Henry86,HamelWoerdman,HW2}.  The Petermann factor is written as
\begin{equation}
K = \left|\frac{\int dr \; |\varphi(r)|^2}{\int dr \;
  \varphi(r)^2}\right|^2 > 1,
  \label{petermann}
\end{equation}
where $\varphi(r)$ is either a transverse or longitudinal wavefunction
(mode amplitude), and the integral is, correspondingly, either taken
over the area transverse to the axis of the laser cavity or along the
axial direction.  The methods which have previously been used to
derive the Petermann factor
\cite{Petermann,Haus,Henry86,Siegman,HamelWoerdman,HW2} are limited to
1D lasers with a well-defined axis, and with spatially uniform
dielectric functions.  In particular, for the longitudinal Petermann
factor, Eq.~(\ref{petermann}) can equivalently be written as
\begin{equation}
  K = \left[\frac{\left(|r_1| +
      |r_2|\right)\left(1-|r_1r_2|\right)}{2|r_1r_2| \, \ln|r_1r_2|}
    \right]^2,
  \label{longitudinal K}
\end{equation}
where $r_1$ and $r_2$ are the reflection coefficients at the two ends
of the uniform 1D cavity \cite{Henry86,HamelWoerdman}.  Recently,
Schomerus has derived a generalization of the Petermann factor which
applies to sub-threshold high-Q 2D lasers \cite{schomerus}.  We will
discuss the relationship between our results and those of Schomerus in
Section \ref{Wavefunction Formulation}.

In this section, we develop a simple and intuitive interpretation of
the Petermann factor, based on the analytic properties of the
$S$-matrix.  In the next section, we will see that the bad-cavity
linewidth reduction factor also emerges in this picture.  The
important role of dispersion in reducing the linewidth was noted in
Ref.~\onlinecite{linewidth_prl}, but in a less complete manner, as we
did not then appreciate its connection to the bad-cavity factor in
Fabry-P\'erot systems.  The geometric interpretation of the Petermann
factor was also touched upon in a recent work on the bandwidth of
coherent perfect absorption \cite{cpanoise}.

\begin{figure}
\centering
\includegraphics[width=0.47\textwidth]{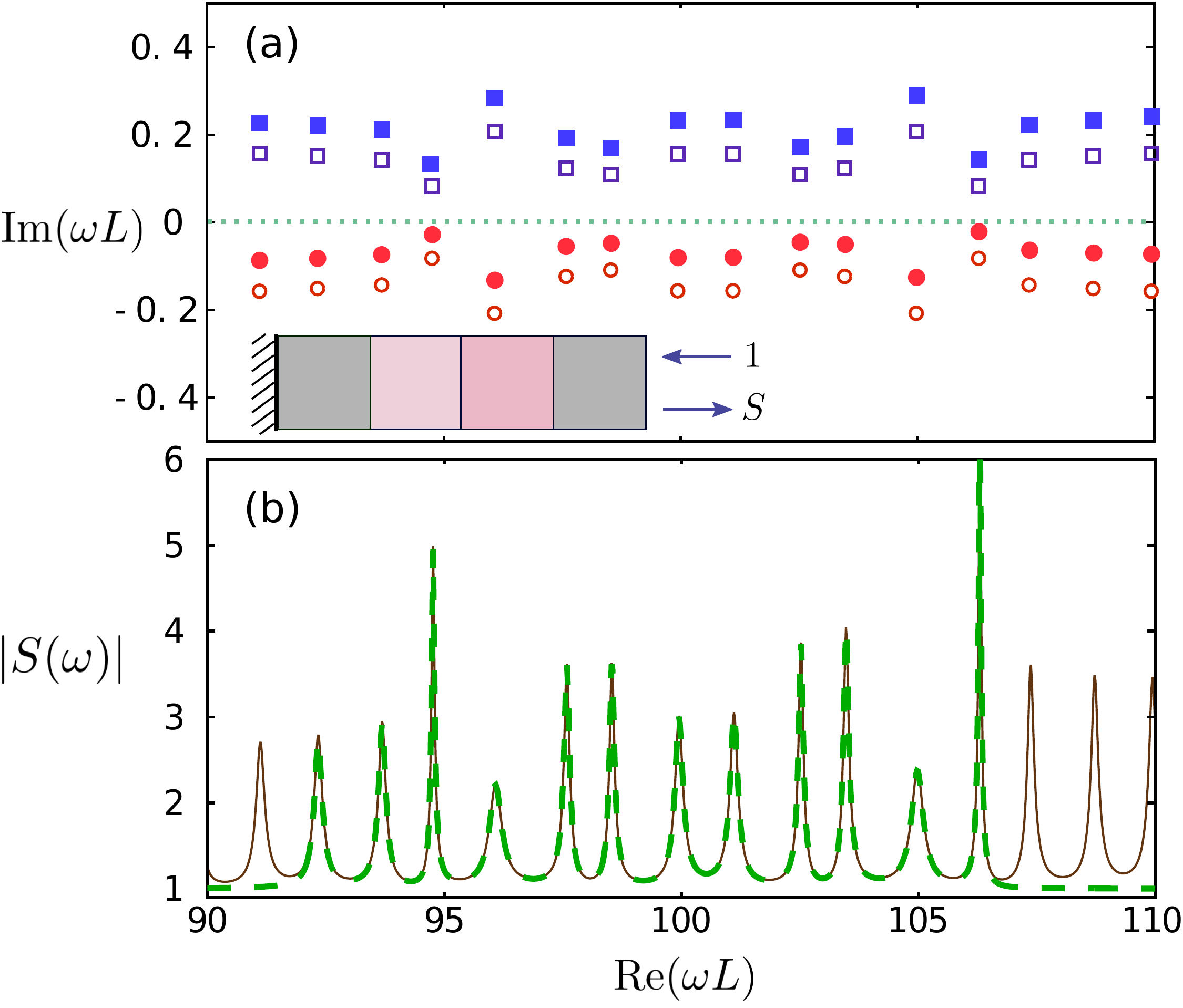}
\caption{(color online) (a) Poles and zeros of a one-port 1D cavity,
  consisting of four slabs of equal length $L/4$ with a perfect mirror
  on the left boundary (inset), and $\epsilon = 1$ in the external
  region.  Filled symbols show poles (circles) and zeros (squares) for
  frequency-independent slab refractive indices of (left to right) $n
  = [3, 1.5-0.002i, 2.5-0.005i, 3]$.  Open symbols show poles and
  zeros for the passive cavity, with $\mathrm{Im}(n) = 0$.  (b) Values
  of $S(\omega)$ on the real-$\omega$ line, for the cavity with gain,
  calculated exactly (solid curve) and using the Pad\'e approximant of
  Eq.~(\ref{pade}) with 12 pole/zero pairs near $\omega L = 100$
  (dashed curve).  }
\label{fig_oneport_omega}
\end{figure}

We begin by considering a one-port laser, such as a 1D cavity with a
perfect mirror on one end, or a higher-dimensional cavity with a
single-mode port.  In this case, $S(\omega)$ is a scalar whose exact
form depends on the distribution of dielectric and gain material in
the system.  Under very general conditions \cite{toll}, $S(\omega)$ is
analytic and possesses an infinite discrete set of poles and zeros in
the complex $\omega$ plane, denoted by $\{\omega^p_j\}$ and
$\{\omega^z_j\}$ respectively.  In a passive cavity (one without
material gain or loss), time-reversal symmetry ensures that the poles
and zeros are symmetrically placed around the real-$\omega$ axis,
allowing us to label them with a single index $j$ such that
$\omega^p_j = (\omega^z_j)^*$.  For example,
Fig.~\ref{fig_oneport_omega}(a) shows the poles and zeros for a
spatially non-uniform 1D cavity (open symbols).  The poles of the
passive cavity correspond to scattering resonances, whose decay rates
(FWHM) are defined as $\gamma_j = -2\,\mathrm{Im}(\omega_j^p)$.

As gain (and/or loss) is introduced, these labeled poles and zeros
move in the complex $\omega$ plane, as shown in
Fig.~\ref{fig_oneport_omega}(a).  They no longer form conjugate pairs,
but if their frequencies $\{\omega^{p,z}_j\}$ are known, we can
compute $S(\omega)$ for any $\omega$ using the Pad\'e approximant
\begin{equation}
  S(\omega) \simeq
  \prod_j \frac{\omega-\omega_j^z}{\omega-\omega_j^p},
  \label{pade}
\end{equation}
up to an irrelevant phase factor which has been omitted.  For the
passive cavity, Eq.~(\ref{pade}) gives $|S(\omega)| = 1$ for all real
$\omega$, as expected.  The precision of the Pad\'e approximant
increases as more pole and zero pairs are included in the product.
Its validity for $S(\omega)$ is demonstrated numerically in
Fig.~\ref{fig_oneport_omega}(b).

A lasing mode corresponds to a pole of $S(\omega)$ located on the
real-$\omega$ axis \cite{spasalt}.  As noted, this description holds
both at the lasing threshold and above threshold, except that
$S(\omega)$ above threshold must be computed using an
$\omega$-dependent dielectric function with a non-linear contribution
from spatial hole-burning, which can be found via the SALT method
\cite{salt1,salt2,salt3,spasalt}.  Denoting the lasing pole by $j =
0$, we can use Eq.~(\ref{generalized decay rate}) with
Eq.~(\ref{pade}) to obtain the generalized decay rate
\begin{equation}
  \gamma_L =
  \left|\omega_0^p-\omega_0^z\right|\,
  \prod_{j\ne 0} \left| \frac{\omega_0^p-\omega_j^z}{\omega_0^p-\omega_j^p}\right|.
  \label{1D ansatz}
\end{equation}
(As mentioned above, for the one-port system, the lasing eigenvector $\Psi$ has only one
component, so $|\Psi^T\Psi| = 1$.)

Fig.~\ref{fig_oneport_omega}(a) demonstrates the effects of broad-band
gain on the poles and zeros.  In this case, the gain is simply
frequency-independent; more generally, ``broad-band'' gain refers to a
gain width much larger than the free spectral range and resonance
decay rates.  The poles and zeros move almost directly upward in the
complex $\omega$ plane, relative to their passive cavity positions,
and each pole and zero in a pair moves by approximately the same
amount.  In order for a given pole (say $j = 0$) to lase, it must move
a distance of $\gamma_0/2$; the corresponding zero moves by the same
amount, so the first factor in Eq.~(\ref{1D ansatz}) is $|\omega_0^p -
\omega_0^z| \approx \gamma_0$.  This factor corresponds to an
unmodified Schawlow-Townes linewidth.

\begin{figure}
\centering
\includegraphics[width=0.32\textwidth]{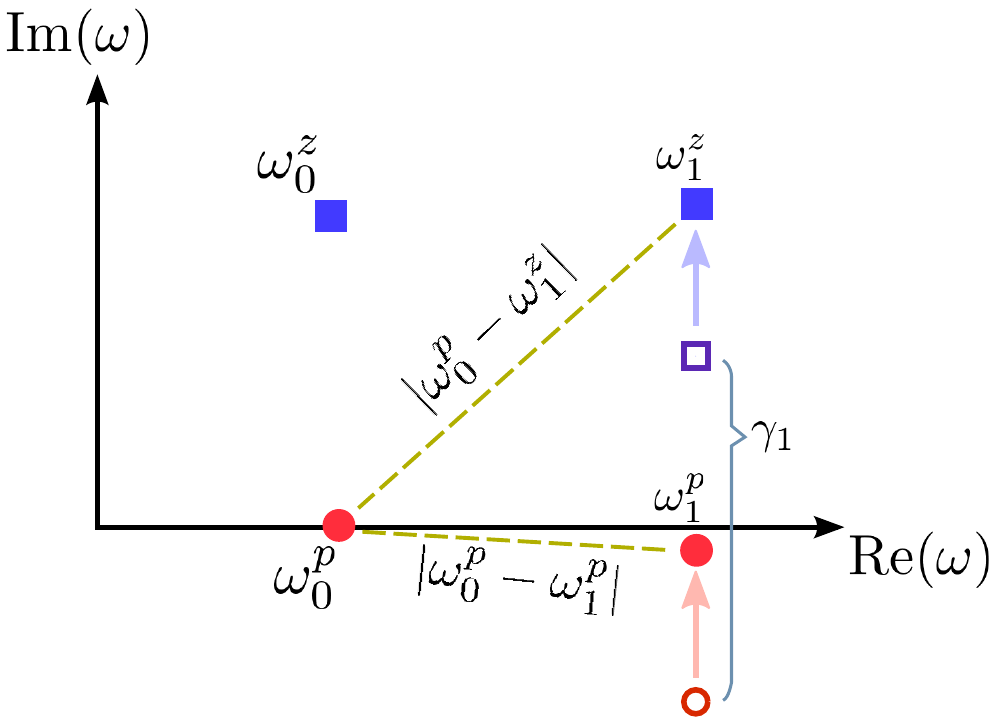}
\caption{(color online) Schematic of a pair of neighboring poles and
  zeros, showing the geometric interpretation of the Petermann factor.
  $\omega_0^p$ is a pole which has reached the real-$\omega$ axis.
  According to Eq.~(\ref{1D ansatz}), the contribution of the
  neighboring pole ($\omega_1^p$) and zero ($\omega_1^z$) to the
  generalized decay rate is the ratio of the lengths of the upper and
  lower dashed lines, which is $> 1$.  Open symbols indicate
  $\omega_1^{p,z}$ for the passive cavity.}
\label{fig_peterman_schematic}
\end{figure}

Next, consider the product terms in Eq.~(\ref{1D ansatz}).
Neighboring pairs of zeros and poles also move upward from their
passive cavity positions.  Hence, as indicated in
Fig.~\ref{fig_peterman_schematic},
\begin{equation}
  K_{\mathrm{ansatz}} \equiv \prod_{j\ne 0} \left|
  \frac{\omega_0^p-\omega_j^z}{\omega_0^p-\omega_j^p}\right|^2 > 1.
  \label{ansatz petermann}
\end{equation}
We interpret $K_{\mathrm{ansatz}}$ as the Petermann factor.  It
approaches unity in the limit where the free spectral range is much
larger than the resonance decay rates $\gamma_j$, in accordance with
the usual notion that the Petermann factor is negligible for high $Q$.

To show that $K_{\mathrm{ansatz}}$ is indeed the Petermann factor,
suppose our one-port laser is spatially uniform and 1D, with
reflection coefficient $r$ at the output port.  The frequencies of the
poles and zeros, denoted $\omega_{p,z|}$, satisfy \cite{cpa}
\begin{equation}
  \exp\left[\pm 2in\omega_{p,z} L\right]\, r = e^{-i\phi},
  \label{Fabry Perot condition 0}
\end{equation}
where $n$ is the refractive index in the cavity, $L$ is the cavity
length, and $\phi$ is the phase change at the perfectly reflecting
port.  For the passive cavity ($\mathrm{Im}[n] = 0$), with
frequency-independent $n$ and $r$, Eq.~(\ref{Fabry Perot condition 0})
implies that the poles and zeroes are equally spaced with free
spectral range $\Delta \omega$ and located at equal distances
$\gamma_0/2$ from the real axis, where $\ln|r| = - \pi\gamma_0/\Delta
\omega$.  Assuming an ``ideal'' gain medium which moves all the poles
up to the real axis, and all the zeros up by an equal amount,
Eq.~(\ref{ansatz petermann}) implies
\begin{equation}
  K_{\mathrm{ansatz}} \simeq \prod_{j\ne 0} \frac{(j\Delta\omega)^2 +
    \gamma_0^2}{(j\Delta\omega)^2}
  = \left[\frac{\Delta \omega}{\pi\gamma_0}
  \, \sinh\left(\frac{\pi\gamma_0}{\Delta \omega}\right)\right]^2.
  \label{fabry K}
\end{equation}
In the last equality, we have used Euler's product formula for the
sine function,
\begin{equation}
  \sin(\pi z) = \pi z \prod_{j=1}^\infty \left(1-\frac{z^2}{j^2}\right),
\end{equation}
with an imaginary argument.  Plugging into Eq.~(\ref{Fabry Perot
  condition 0}) yields
\begin{equation}
  K_{\mathrm{ansatz}} \simeq \left|\frac{1-r^2}{2r\ln |r|}\right|^2.
  \label{woerdman K}
\end{equation}
This agrees exactly with Eq.~(\ref{longitudinal K}), the formula for
the longitudinal Petermann factor derived in in
Refs.~\onlinecite{Henry86,HamelWoerdman}, for the one-port case ($r_1
= r, |r_2| = 1$).  This link between the motion of $S$-matrix poles
and zeros and the Eq.~(\ref{longitudinal K}) is a new result of this
paper.  This geometric interpretation also emphasizes the fact that
the Petermann factor relates to the cavity finesse, $\Delta \omega /
\gamma_0$, not the $Q$-factor $\omega_0 / \gamma_0$.

For a cavity with more than one port, the ansatz (\ref{1D ansatz}) no
longer applies since the $S$-matrix has more than one eigenvalue.  (As
an exception, in a two-port parity symmetric system, the eigenspace of
$S$ factorizes and Eq.~(\ref{1D ansatz}) can be used with only
even/odd values of $j$ in the product.)  In the more general case,
$\gamma_L$ would have to be calculated using Eq.~(\ref{generalized
  decay rate}), or from the wavefunction formula derived in Section
\ref{Wavefunction Formulation}.

\section{The bad-cavity factor}
\label{bad cavity section}

In the previous section, when showing that the Pad\'e approximant
ansatz (\ref{1D ansatz}) for the generalized decay rate yields the
Schawlow-Townes-Petermann linewidth for broad-band gain, we assumed
that the gain displaces the poles from their passive cavity positions
by the same amount as the zeros.  Thus, for instance, the leading
factor of $|\omega_0^p - \omega_0^z|$ in Eq.~(\ref{1D ansatz}) takes
the value $\gamma_0$.  When the dielectric function is frequency
dependent, this condition is violated.  Consider a Maxwell-Bloch gain
medium,
\begin{equation}
  \epsilon(\omega) = n_0^2
  + \frac{D\gamma_\perp}{\omega - \omega_a + i\gamma_\perp},
  \label{maxwell bloch eps}
\end{equation}
where $n_0^2$ is the background permittivity, $D$ a scaled inversion
factor proportional to the pump, $\omega_a$ the polarization resonance
frequency, and $\gamma_\perp$ the polarization dephasing rate (gain
width) \cite{salt1,salt2,salt3,spasalt}.  This formula for
$\epsilon(\omega)$ can be analytically continued into the complex
$\omega$ plane \cite{toll}, in order to compute $S(\omega)$ for
complex $\omega$.

\begin{figure}
\centering
\includegraphics[width=0.45\textwidth]{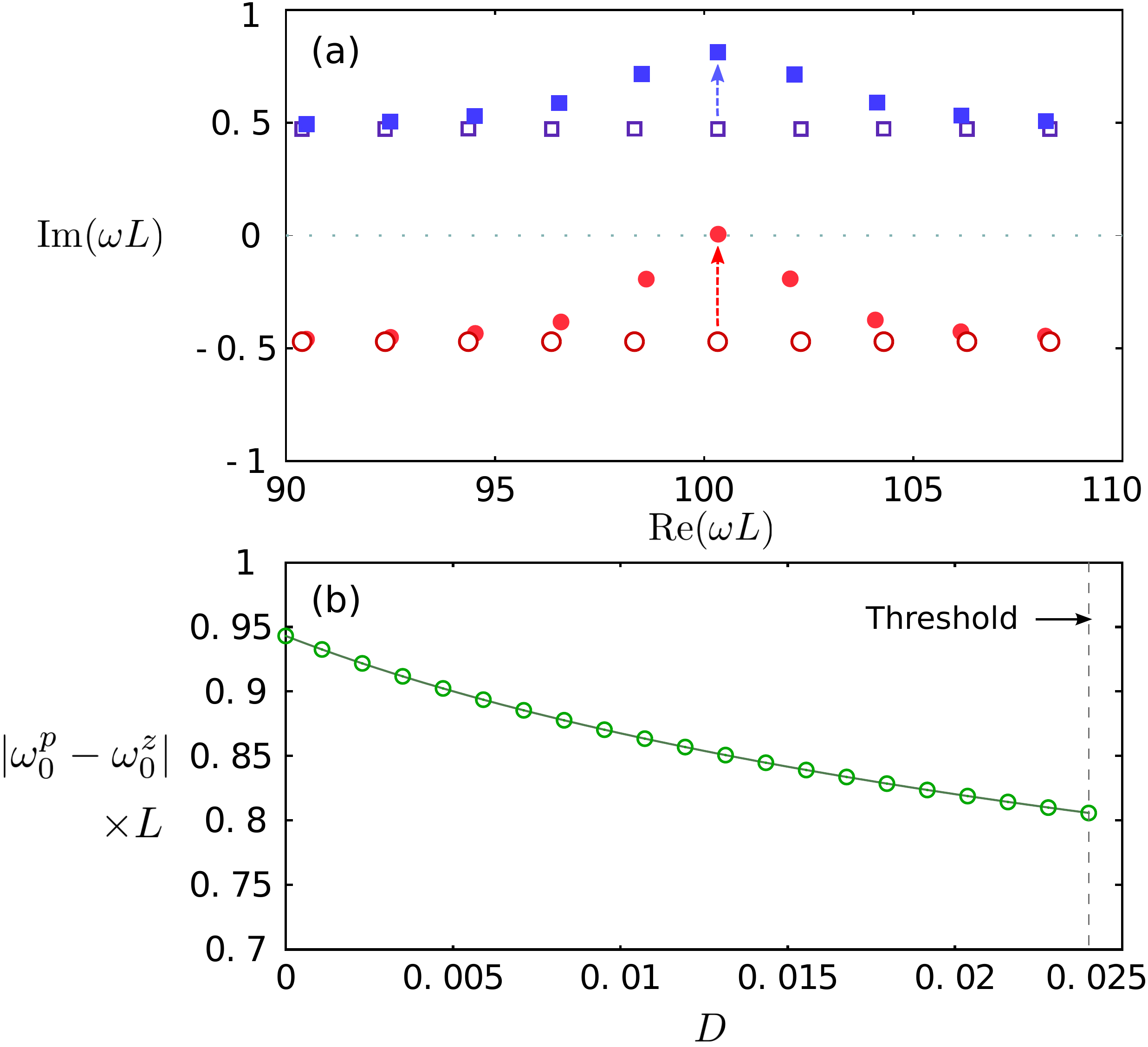}
\caption{(color online) (a) Effect of Maxwell-Bloch gain medium on
  poles and zeros.  Poles (circles) and zeros (squares) are plotted
  for a one-port Fabry-P\'erot cavity of length $L$, for passive
  dielectric $\epsilon = 2.5$ (open symbols) and for a Maxwell-Bloch
  medium (filled symbols) with $n_0^2 = 2.5$, $\gamma_\perp = 2/L$,
  $\omega_a = 100.3/L$, and $D = 0.024$.  The gain medium moves the
  poles up by a greater distance than the zeros, resulting in the gain
  dispersion linewidth correction factor of
  Ref.~\onlinecite{Woerdman}.  (b) Values of $|\omega_0^p -
  \omega_0^z|$ for the zero and pole closest to the gain center,
  versus the pump $D$.  The dots are exact numerical solutions; the
  line is an approximation using Eq.~(\ref{gammapz}). }
\label{fig_mb_poles}
\end{figure}

Fig.~\ref{fig_mb_poles}(a) shows the poles and zeros for a one-port
Fabry-P\'erot cavity with this dielectric function.  Both the poles
and zeros are moved upward, \textit{but the zeros move by a smaller
  distance}.  As a result, in Eq.~(\ref{1D ansatz}) the leading factor
of $|\omega_0^p-\omega_0^z|$ is smaller than $\gamma_0$, and the
product terms (which, as discussed above, give rise to the Petermann
factor) are likewise reduced.

The fact that the zeros move less than the poles can be understood
intuitively from Eq.~(\ref{maxwell bloch eps}).  The effect of the
gain medium (which pushes poles and zeros upward in the complex plane)
is large when $\omega$ is close to $\omega_a - i\gamma_\perp$, which
lies in the lower half-plane.  Hence the zeros, which are in the upper
half plane, ``experience less gain'' than the poles in the lower half
plane. The resulting linewidth reduction can, in principle, overcome
the increase due to the Petermann factor, resulting in a linewidth
\textit{below} the Schawlow-Townes limit.

The linewidth reduction can be quantified in a simple way for our toy
model of a spatially uniform 1D one-port cavity, for which the pole
and zero frequencies $\omega_{p,z}$ are given by Eq.~(\ref{Fabry Perot
  condition 0}).  For the passive cavity, $n = n_0$, suppose that
there is a pair of poles and zeros located at $\omega_0 \mp i \gamma_0
/ 2$.  For the pumped cavity, the refractive index is $n = n' + i n''$
and the central pole and zero frequencies become $\omega_{p,z} =
\omega_{p,z}' + i \gamma_{p,z}$.  If $r$ is approximately independent of $n$ and
$\omega$, Eq.~(\ref{Fabry Perot condition 0}) gives
\begin{eqnarray}
  \omega_{p,z}' &=& \frac{1}{|n|^2} \left[\mp n_0 n'' \frac{\gamma_0}{2} + n_0 n' \omega_0 \right] \\
  \gamma_{p,z} &=& \frac{1}{|n|^2} \left[\mp n_0 n' \frac{\gamma_0}{2} - n_0 n'' \omega_0\right].
  \label{gammapz}
\end{eqnarray}
Threshold occurs when $n''/n' = \gamma_0 / 2\omega_0$.  For a high-Q
cavity ($\gamma_0 \ll \omega_0$), this implies $|n''| \ll |n'|$ and
$\omega_{p,z}' \approx \omega_0$, as expected.  Now suppose the medium
has the Maxwell-Bloch form, with the gain curve centered on this pair
of poles and zeros ($\omega_a = \omega_0$).  From Eq.~(\ref{maxwell
  bloch eps}), at the pole and zero frequencies,
\begin{equation}
  \frac{n''}{n'} \approx - \frac{D\gamma_\perp}{2n_0^2\,(\gamma_{p,z} +
    \gamma_\perp)}. \label{npp}
\end{equation}
Threshold occurs at $D = n_0^2 \gamma_0 / \omega_0$.  From
Eqs.~(\ref{gammapz})-(\ref{npp}), we can find the imaginary part of
the zero frequency at threshold:
\begin{eqnarray}
  \gamma_z &=& \frac{\gamma_0}{2} \left(1 +
    \frac{\gamma_\perp}{\gamma_z + \gamma_\perp}\right) \\ &=&
    \gamma_0 \, \left[1 + \frac{\gamma_z}{2\gamma_\perp} \,
      \left(1+\frac{\gamma_z}{2\gamma_\perp}\right)^{-1} \right]^{-1}.
\end{eqnarray}
Hence,
\begin{equation}
  \gamma_z \ge \gamma_0 \, \left[1 + \frac{\gamma_0}{2\gamma_\perp}
    \right]^{-1},
  \label{ansatz bad cavity}
\end{equation}
with the inequality saturating as $\gamma_0 \ll \gamma_\perp$.  This
result is valid for a high-Q cavity at threshold.

The right hand side of (\ref{ansatz bad cavity}) is $\gamma_0$
multiplied by a factor smaller than unity, which is precisely the
``bad-cavity'' factor previously derived in
Refs.~\onlinecite{Lax,Haken,Kolobov,Woerdman}.  For the moment, let us
consider the perturbative limit, $\gamma_0 \ll \gamma_\perp$, where
this factor is comparable to unity and the inequality (\ref{ansatz bad
  cavity}) saturates.

For a high-finesse cavity ($\gamma_0 \ll \Delta \omega$, occurring for
$|r| \rightarrow1$), the product terms in the ansatz (\ref{1D ansatz})
go to unity, so $\gamma_L \approx \gamma_z$.  In this limit,
Eq.~(\ref{ansatz bad cavity}) gives rise to a laser linewidth which
includes a ``bad-cavity'' factor and a negligible Petermann factor, as expected.

\begin{figure}
\centering
\includegraphics[width=0.45\textwidth]{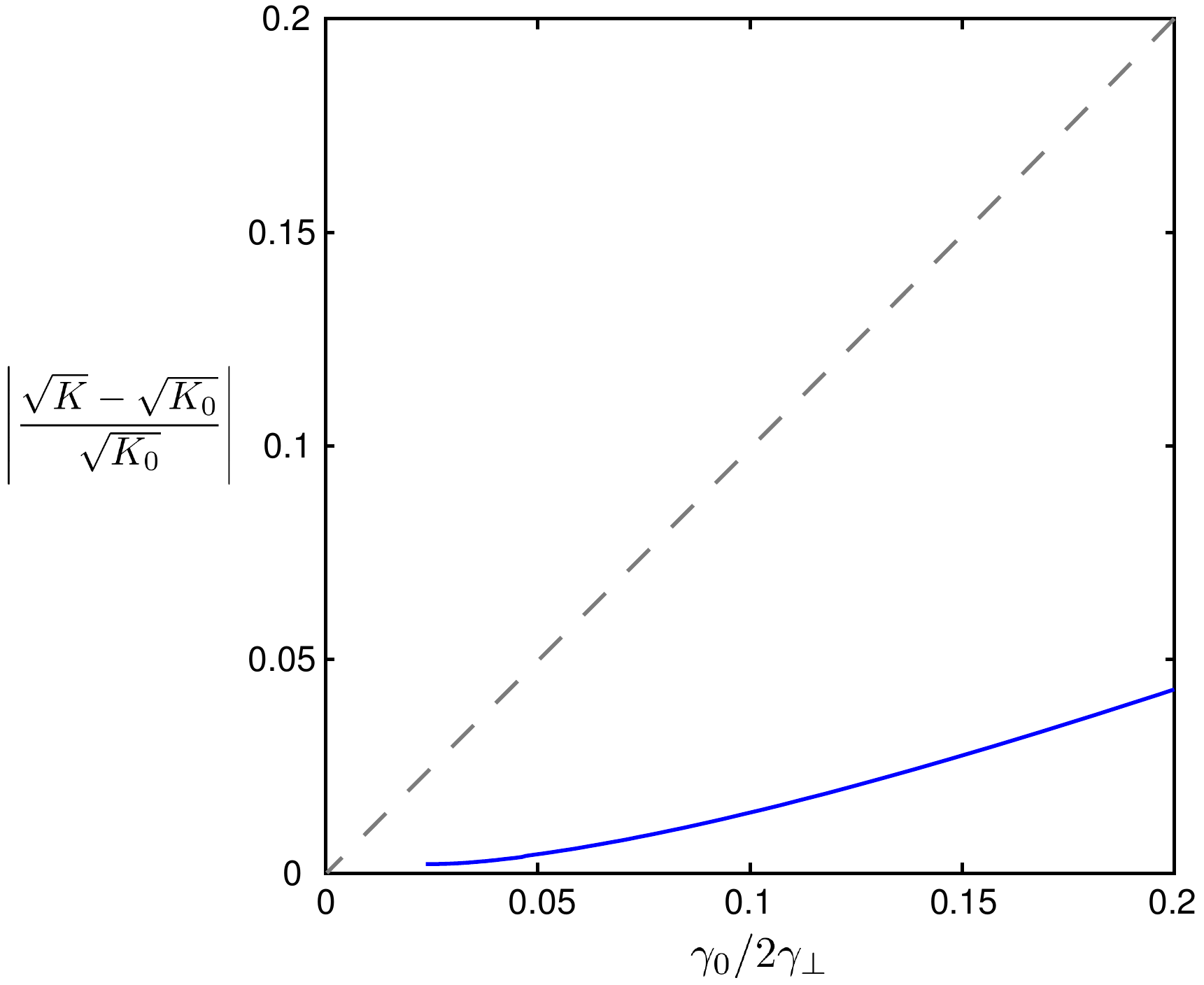}
\caption{(color online) Fractional deviation in $\sqrt{K}$ caused by
  finite Maxwell-Bloch gain width $\gamma_\perp$, as a function of
  $\gamma_0/2\gamma_\perp$ (where $\gamma_0$ is the passive cavity
  decay rate).  This plot is obtained varying $\gamma_\perp$ in the
  one-port Fabry-P\'erot cavity of Fig.~\ref{fig_mb_poles}, and
  calculating the Petermann factor from Eq.~(\ref{ansatz petermann}),
  based on the pole and zero frequencies at threshold.  The broad-band
  limit of the Petermann factor, $K_0$, is computed from
  Eq.~(\ref{fabry K}). }
\label{K_deviation}
\end{figure}

Away from the high-finesse limit ($\gamma_0 \sim \Delta \omega$), the situation is 
less clear and the Petermann factor cannot be neglected.  In the ansatz (\ref{1D
  ansatz}), we must account for the other pairs of poles and zeros,
which are located away from the gain center $\omega_a$.  For example,
Fig.~\ref{fig_mb_poles}(a) shows the poles and zeros for a
medium-finesse Fabry-P\'erot cavity with $\gamma_0 \approx 0.5
\Delta\omega$.  The non-central poles and zeros are displaced upwards
from their passive-cavity frequencies, but by less than the central
pair.  They also experience ``line-pulling'' towards the gain center
$\omega_a$.  Both effects modify the Petermann factor. The first tends to suppress $K$,
while the second tends to enhance it. Let us compare
$\sqrt{K_{\mathrm{ansatz}}}$, as calculated from Eq.~(\ref{ansatz
  petermann}) with the Maxwell-Bloch gain medium, to the value
$\sqrt{K_0}$ for ``infinitely broad-band'' gain, which is given by
Eq.~(\ref{fabry K}).  Fig.~\ref{K_deviation} shows the value of
\begin{equation*}
  \left|\frac{\sqrt{K_{\mathrm{ansatz}}} -
    \sqrt{K_0}}{\sqrt{K_0}}\right|
\end{equation*}
versus $\gamma_0/2\gamma_\perp$, as $\gamma_\perp$ is varied in the
Fabry-P\'erot cavity.  It turns out that the effects of gain reduction and line pulling on
the Petermann factor cancel to first order leaving a correction which is
second-order in $\gamma_0/2\gamma_\perp$.  Hence, even if the
Petermann factor itself is non-negligible, the correction to it due to a
finite $\gamma_\perp$ are negligible compared to the correction to
$\gamma_z$.  The bad-cavity factor and the Petermann factor can be
taken to be independent, in the limit $\gamma_0 \ll \gamma_\perp$.

Away from the $\gamma_0 \ll \gamma_\perp$ limit, the inequality
(\ref{ansatz bad cavity}) does not saturate, and the ``bad-cavity''
linewidth reduction effect has a significant contribution from the
positions of the non-central poles and zeros.  In this case, the
bad-cavity and Petermann factor cannot be cleanly identified with the
leading and product terms in the ansatz (\ref{1D ansatz}).  In the
next section, we will derive a general linewidth formula in terms of
the lasing wavefunction, which is valid even if $\gamma_0 \sim
\gamma_\perp$.  We will then show that the bad-cavity correction
appears as a separate multiplicative factor, consistent with
Refs.~\onlinecite{Lax,Haken,Kolobov,Woerdman}, only for the special
case of a Fabry-P\'erot cavity. In general the Petermann and
bad-cavity factors cannot be separated when $\gamma_0 \gtrsim
\gamma_\perp$, i.e.~in the full bad-cavity regime.

\section{Wavefunction Formulation}
\label{Wavefunction Formulation}

The generalized decay rate $\gamma_L$, defined in
Eq.~(\ref{generalized decay rate}), can be re-expressed in terms of
the wavefunction of the lasing mode and the frequency-dependent
dielectric function of the laser medium.  This derivation, which uses
a modification of the approximation scheme of Schomerus
\cite{schomerus}, yields a generalization of the usual Petermann
formula (\ref{petermann}) that also incorporates the bad-cavity
factor.  Pick \textit{et al.} have independently derived a similar
result, using a coupled mode theory approach which also yields a
generalization of the Henry $\alpha$ factor \cite{pick}.

Consider a 1D or 2D transverse magnetic (TM) system lasing at a real
frequency $\omega = \omega_0$.  The laser mode is described by a
purely-outgoing wavefunction $\psi_0$ (representing the out-of-plane
component of the complex electric field), which satisfies the
Helmholtz equation:
\begin{align}
  \begin{aligned}
    \Big[\nabla^2 + \epsilon(r,\omega_0) &\,\omega_0^2\, \Big]
    \psi_0(r) = 0,\\
    \psi_0(r) &= \sum_\mu b_\mu u_\mu(r;\omega_0)\;\;\mathrm{for}\;\;
    r \notin C.
    \label{lasing mode}
  \end{aligned}
\end{align}
Here, $C$ denotes the scattering region, and $\{u_\mu\}$ is an
appropriate set of outgoing channel modes defined in the region
outside $C$, where $\epsilon = 1$.  For open 2D geometries, it is
convenient to let $C$ be a circle of radius $R$, and define
\begin{equation}
  u_\mu(r,\phi;\omega) = \frac{H^{+}_\mu(\omega
    r)}{\sqrt{R}\,H^{+}_\mu(\omega R)} \Phi_\mu(\phi),
  \label{2d channel modes}
\end{equation}
with azimuthal basis functions satisfying $\int_0^{2\pi} d\phi\,
\Phi_\mu\Phi_\nu = \delta_{\mu\nu}$.  The vector $\mathbf{b} = [b_1,
  \cdots]$ is an eigenvector of $S(\omega_0)$, with diverging
eigenvalue.

Next, consider a frequency $\omega$ differing slightly from
$\omega_0$.  The $S$-matrix remains dominated by the pole, so
\cite{schomerus}
\begin{equation}
  S(\omega) \approx \frac{\sigma(\omega)}{\mathbf{b}^T \mathbf{b}}
  \; \mathbf{b}\mathbf{b}^T,
  \label{Sapprox}
\end{equation}
with $\sigma(\omega)$ finite.  Let $\mathbf{a}$ be an input amplitude,
normalized so that the output amplitude is equal to $\mathbf{b}$.
From (\ref{Sapprox}), the generalized decay rate is
\begin{equation}
  \gamma_L = \left|\mathrm{Res}
  \left[\sigma(\omega) \, \frac{\mathbf{b}^\dagger
      \mathbf{b}}{\mathbf{b}^T \mathbf{b}}
    \right]\right| = 
  \mathbf{b}^\dagger
  \mathbf{b}
  \left|\,\mathrm{Res}\left(\frac{1}{\mathbf{b}^T \mathbf{a}}\right)\right|,
  \label{gammal_approx}
\end{equation}
with the residue evaluated in the limit $\omega\rightarrow\omega_0$.
The corresponding wavefunction, $\psi(r)$, obeys
\begin{align}
  \begin{aligned}
    \Big[\nabla^2 \,+ \,&\epsilon(r,\omega) \,\omega^2\, \Big]
    \psi(r) = 0,\\
    \psi(r) &= 
    \sum_\mu \left[
    a_\mu u_\mu^*(r) +
    b_\mu u_\mu(r)\right]\;\mathrm{for}\;\;
    r \notin C.
    \label{wavefunction}
  \end{aligned}
\end{align}
According to Gauss's theorem,
\begin{align}
  \begin{aligned}
    \int_C d^dr &\, \Big[\psi_0 \nabla^2 \psi - \psi \nabla^2 \psi_0\Big]
  = \Big[\psi_0\, \nabla \psi
  - \psi\, \nabla \psi_0\Big]_{\partial C} \\
  &= -i(\omega+\omega_0) \mathbf{b}^T \mathbf{a} + i(\omega-\omega_0) \mathbf{b}^T \mathbf{b}.
    \label{integral by parts}
  \end{aligned}
\end{align}
The final equality in (\ref{integral by parts}) is exact for 1D, and
approximate for the 2D modes defined in (\ref{2d channel modes}) in
the limit $\omega R \gg 1$.  From the wave equation, (\ref{integral by
  parts}) also equals
\begin{multline}
    \int_C d^dr\, \Big[\epsilon(r,\omega_0) \omega_0^2
    - \epsilon(r,\omega)\omega^2\Big] \psi_0\psi \\
  \approx - (\omega - \omega_0)
  \int_C d^dr \left[\omega^2\frac{d\epsilon}{d\omega}
    + 2\epsilon\,\omega \right]_{\omega_0} \psi_0^2.
  \label{schomerus approx 2}
\end{multline}
Exploiting the time reversal symmetry of the Helmholtz equation,
$\psi_0^*$ acts as a purely incoming solution with $\epsilon \rightarrow
\epsilon^*$. This gives
\begin{align}
  \begin{aligned}
    \int_C d^dr &\, \Big[\psi_0 \nabla^2 \psi^*_0 - \psi^*_0 \nabla^2 \psi_0\Big]
  = \Big[\psi_0\, \nabla \psi^*_0
  - \psi^*_0\, \nabla \psi_0\Big]_{\partial C} \\
  &= -i(\omega^*_0+\omega_0) \mathbf{b}^{\dagger} \mathbf{b},
    \label{integral by parts conjugate}
  \end{aligned}
\end{align}
and using the wave equation as before,
\begin{align}
  \begin{aligned}
	\int_C d^dr\, &\Big[\epsilon(r,\omega_0) \omega_0^2
	- \epsilon^*(r,\omega_0)(\omega_0^*)^2\Big] |\psi_0|^2 \\
	&= \int_C d^dr 2\mathrm{Im} [\epsilon(r,\omega) \omega_0^2]
	|\psi_0|^2.
	\label{schomerus approx conjugate}
  \end{aligned}
\end{align}
Combining these equations and using (\ref{gammal_approx}) yields
\begin{equation}
  \gamma_L = \frac{\omega_0 \displaystyle \int_C d^dr\,
  \mathrm{Im}[\epsilon(r,\omega_0) \omega_0^2]
    \; |\psi_0|^2 }{\mathrm{Re}(\omega_0) \displaystyle\left| 
    \frac{i\mathbf{b}^T \mathbf{b}}{2}
    + \int_C d^dr \left[
      \epsilon\omega + \frac{\omega^2}{2}\frac{d\epsilon}{d\omega} \right]_{\omega_0}
    \psi_0^2\;
     \right|}.
     \label{gammal_integral complex freq}
\end{equation}

Eq.~(\ref{gammal_integral complex freq}) expresses the generalized cavity decay
rate in terms of the lasing mode (\ref{lasing mode}), valid for
arbitrary cavity geometries and gain media.  We now show that it
reduces to the usual bad-cavity linewidth formula for the special
case of a uniform 1D Fabry-P\'erot cavity of length $L$. The
$\mathbf{b}^T \mathbf{b}/2$ term in the denominator is normalized
to the value of $\psi(r)$ at the cavity boundary, in accordance with
Eq.~(\ref{wavefunction}). We will drop this term, as it is negligible
for $\omega_0 L \gg 1$. For $\omega_0 = \mathrm{Re}(\omega_0)$, Eq.~(\ref{gammal_integral complex freq})
becomes
\begin{equation}
  \gamma_L = B \, \sqrt{K} \, \gamma_0,
  \label{Fabry-Perot gamma form}
\end{equation}
where
\begin{eqnarray}
  B &=& \left|1+\frac{\omega}{2\epsilon}\,
  \frac{d\epsilon}{d\omega} \right|_{\omega_0}^{-1} \label{frequency beta} \\
  K &=& \left|\frac{\int dz\; |\psi_0|^2 }{\int dz\; \psi_0^2}\right|^2 \\
  \gamma_0 &=& - \frac{\omega_0\,
    \mathrm{Im}[\epsilon(\omega_0)]}{|\epsilon(\omega_0)|}
  \label{gamma0}
\end{eqnarray}
are respectively the bad-cavity factor, the longitudinal Petermann
factor \cite{HamelWoerdman}, and the passive cavity decay width.  In
(\ref{gamma0}), $\gamma_0$ was identified via the standard relation
between the passive cavity resonance frequency $\omega_c$ and the
lasing refractive index $n_0$ in a Fabry-P\'erot cavity \cite{cpa}:
\begin{equation}
  \frac{\mathrm{Im}(n_0)}{\mathrm{Re}(n_0)} = 
  \frac{\mathrm{Im}(\epsilon)}{2\mathrm{Re}(\epsilon)} \approx
  \frac{\mathrm{Im}(\omega_c)}{\mathrm{Re}(\omega_c)} \approx
  -\frac{\gamma_c/2}{\omega_0}.
  \label{fabry perot}
\end{equation}
This approximation assumes that line-pulling is negligible. We now
focus on the $B$ factor of Eq.~(\ref{frequency beta}).  Using the
Maxwell-Bloch dielectric function (\ref{maxwell bloch eps}) in the
limit $\omega_a \approx \omega_0 \gg \gamma_\perp$, together with
Eq.~(\ref{fabry perot}), Eq.~(\ref{frequency beta}) simplifies to
\begin{eqnarray}
  B \approx \left|1+\frac{\omega_0 D}{2n_0^2\gamma_\perp}
  \right|^{-1} \approx \left|1+\frac{\gamma_c}{2\gamma_\perp}
  \right|^{-1}.
\end{eqnarray}

\begin{figure}
\centering \includegraphics[width=0.43\textwidth]{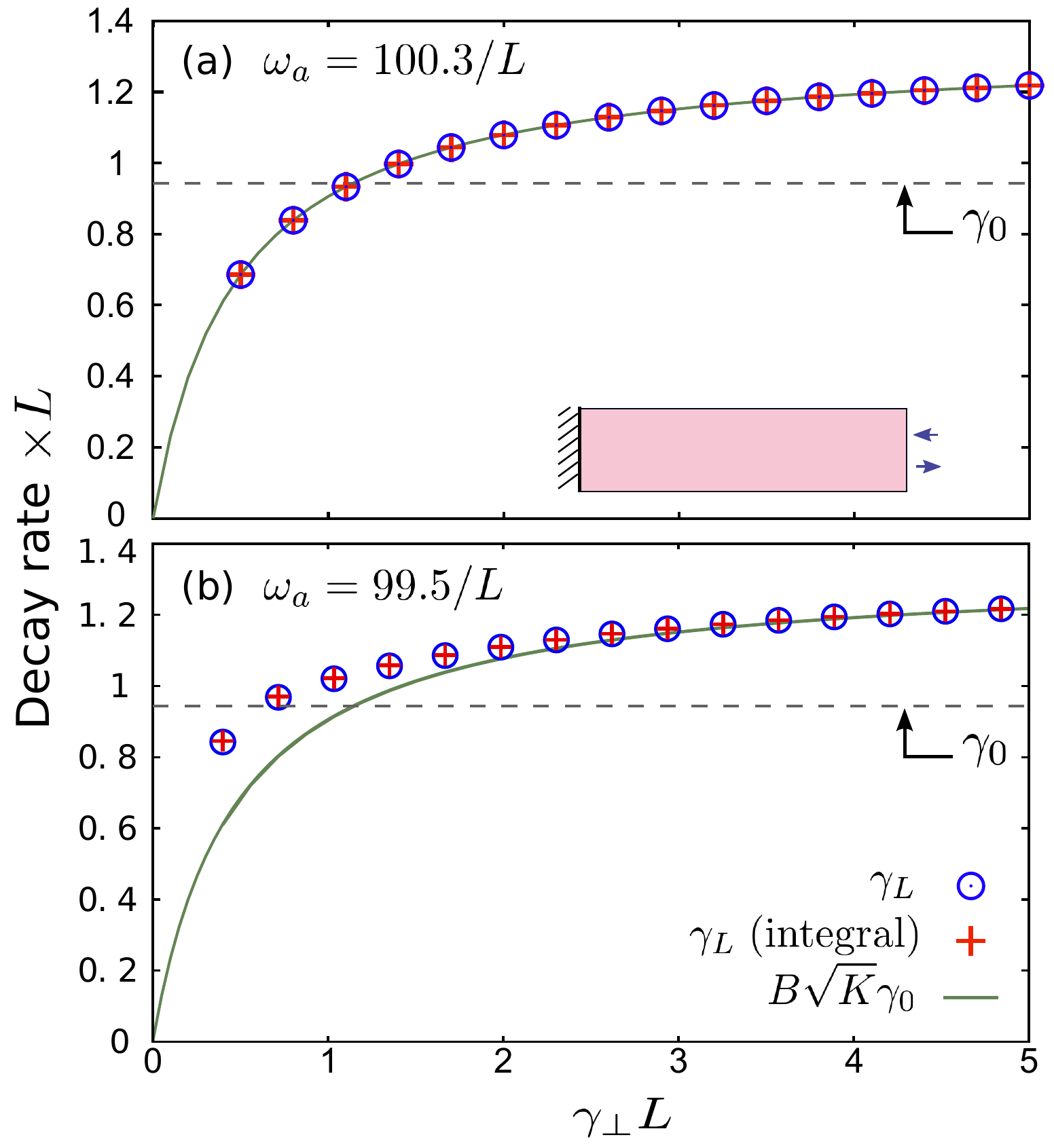}
\caption{(color online) Generalized decay rates for a one-port
  Fabry-P\'erot laser at threshold, versus the gain width
  $\gamma_\perp$.  The Maxwell-Bloch gain medium has background
  permittivity $n_0^2 = 2.5$, with the pump adjusted so that the
  laser is at threshold.  Results are shown for (a) $\omega_a =
  100.3/L$, centered on one of the poles of the passive cavity; and
  (b) $\omega_a = 99.5/L$, detuned from the pole by slightly less than
  a free spectral range.  In both plots, the decay rates are
  calculated using the exact formula for $\gamma_L$ from
  Eq.~(\ref{generalized decay rate}) (blue circles), using the
  integral approximation (\ref{gammal_integral complex freq}) (red crosses),
  and using the Fabry-P\'erot-specific Eqs.~(\ref{Fabry-Perot gamma
    form})-(\ref{gamma0}) (solid line).  The passive cavity decay rate
  $\gamma_0$ is also shown (dashed line). }
\label{fig_linewidth_gperp}
\end{figure}

The above results are verified numerically in
Fig.~\ref{fig_linewidth_gperp}, for one-port Fabry-P\'erot lasers at
threshold (with varying $\gamma_\perp$).  The $S$-matrix theory is
typically in good agreement with the Fabry-P\'erot specific
Eqs.~(\ref{Fabry-Perot gamma form})-(\ref{gamma0}).  Discrepancies
are, however, observed when $\omega_a$ is significantly detuned from a
lasing pole and $\gamma_\perp$ is small, as shown in
Fig.~\ref{fig_linewidth_gperp}(b).  In this regime, the lasing modes
are strongly affected by line-pulling, so that the factorization
(\ref{Fabry-Perot gamma form}) breaks down.  In all the studies we
have performed, the integral form of the generalized decay rate,
Eq.~(\ref{gammal_integral complex freq}), is in excellent agreement
with $\gamma_L$ as evaluated directly from the $S$-matrix. It should
be noted that both approaches require solving the SALT equations for
the (in general) non-linear multimode lasing state, but once that is
done the integral formula is evaluated simply by performing the
relevant integrals of SALT solutions over the lasing cavity, whereas
the S-matrix residue formula typically requires more involved
calculations.

The key approximations (\ref{Sapprox}) and (\ref{integral by
  parts})-(\ref{schomerus approx conjugate}) were previously used by Schomerus
to derive a generalization of the Petermann factor for 2D lasers
\cite{schomerus}.  The method of Schomerus differs from ours in
several respects.  Instead of calculating the $S$-matrix eigenvalue
residue, he calculated the amplified spontaneous emission (ASE)
intensity,
\begin{equation}
  I(\omega) \approx \frac{1}{2\pi} \mathrm{Tr}(S^\dagger S)
  \approx \frac{1}{2\pi} \left|\frac{\mathbf{b}^\dagger \mathbf{b}}
          {\mathbf{b}^T \mathbf{a}}\right|^2,
   \label{ASE}
\end{equation}
for a \textit{sub-threshold} laser cavity.  In Eqs.~(\ref{integral by
  parts})-(\ref{schomerus approx 2}), the wavefunction $\psi$ is
chosen to be that of the sub-threshold system, whose dielectric
function $\epsilon$ differs from the threshold laser's dielectric function
$\epsilon_L$.  For real $\omega$, Eq.~(\ref{ASE}) gives a Lorentzian
with width $\Delta \omega$, inverse to the total ASE power $P = \hbar
\omega_0 \int I(\omega)\,d\omega$; this Schawlow-Townes-like
relationship is argued to hold as the system approaches threshold, at
which point it becomes the laser linewidth.  Furthermore, $I(\omega)$
diverges at a complex frequency below the real-$\omega$ axis,
corresponding to the pole of the sub-threshold cavity.  By assuming
that Eq.~(\ref{ASE}) holds all the way down to the passive cavity
limit, where the dielectric function is
$\approx\mathrm{Re}(\epsilon_L)$ and the resonance frequency is
$\approx \omega_0-i\gamma_0/2$, Schomerus obtains \cite{schomerus}
\begin{equation}
  \Delta \omega \approx \frac{\hbar \omega_0 \gamma_0^2}{P_{\mathrm{tot}}}
  \left|\frac{\int d^2r\; \mathrm{Im}(\epsilon_L)\, |\psi_0|^2 }{\int d^2r \; \mathrm{Im}(\epsilon_L)\, \psi_0^2}\right|^2.
  \label{Schomerus}
\end{equation}
Note that this reduces to the traditional formula for the longitudinal
Petermann factor for uniform cavities.

Thus, in Schomerus' theory the relevant approximations are based on a
perturbation between a passive cavity pole and a lasing pole.  By
contrast, we have used the approximation (\ref{schomerus approx 2}) to
describe a truly infinitesimal deviation from an $S$-matrix pole, for
the purpose of extracting the residue of the pole.  As might be
expected, the Schomerus result (\ref{Schomerus}) agrees well with our
present theory when $\gamma_0$ is much smaller than all other
frequency scales, including $\gamma_\perp$ and the free spectral
range.  However the two do not agree in other regimes, when dispersion
is non-negligible, since the frequency dependence of $\epsilon$ is
ignored.  In the next section we will correct the Schomerus theory by
including the traditional bad-cavity factor (\ref{ansatz bad cavity})
by hand and use this hybrid theory to compare to our more complete
theory for complex cavities.
%We will find that the hybrid theory is often adequate but that
%line-pulling and spatial hole-burning can lead to significant
%interactions between the bad-cavity and generalized Petermann effects,
%so that the more complete $S$ matrix theory must be employed. DO WE NEED this?

\section{Complex laser cavities}
\label{numerical section}

Having established that the $S$-matrix theory agrees analytically and
numerically with previous theories for uniform 1D lasers, we now turn
to more complex cases---1D lasers with spatially non-uniform
dielectric functions and pumping, 2D lasers, and the effect of spatial
hole-burning above threshold.

\begin{figure}
\centering \includegraphics[width=0.46\textwidth]{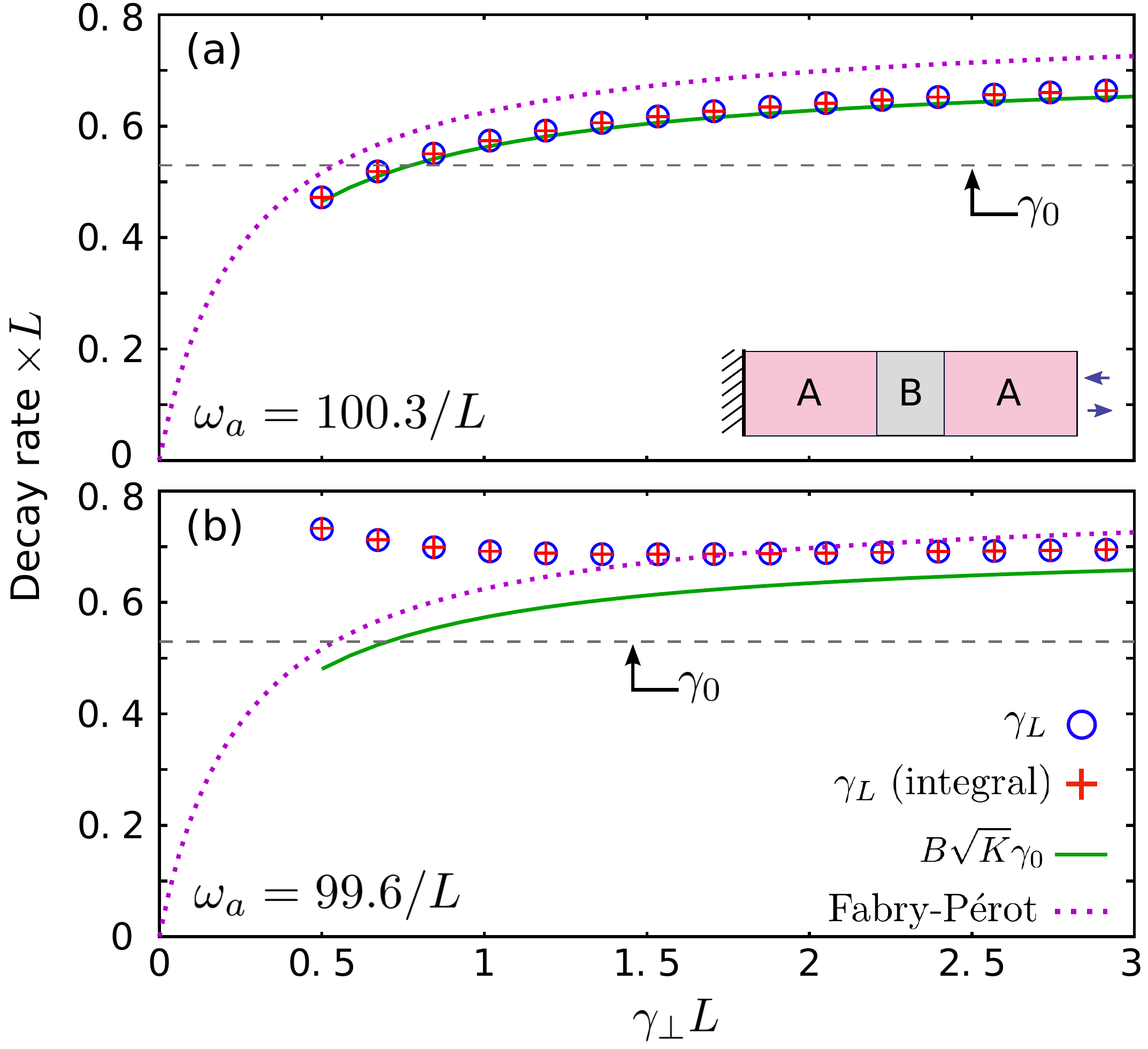}
\caption{(color online) Generalized decay rates versus gain width
  $\gamma_\perp$, for a partially-pumped non-uniform laser at
  threshold with the gain centers (a) $\omega_a = 100.3/L$ and (b)
  $\omega_a = 99.6/L$.  The laser consists of three slabs of lengths
  $[0.4L,0.2L,0.4L]$ (inset schematic).  The A slabs contain gain
  material with background $n_0^2 = 2.25$; the B slab has passive
  $\epsilon = 9$.  The generalized decay rate $\gamma_L$ is computed
  from the $S$-matrix (blue circles), and from the integral
  approximation Eq.~(\ref{gammal_integral complex freq})
  (red crosses).  The solid curves show the traditional result
  $B \sqrt{K} \gamma_c$, where $B$ is the bad-cavity factor of
  Eq.~(\ref{frequency beta}) and $K$ is the Petermann factor of
  Eq.~(\ref{Schomerus}).  The dotted curves show the result for
  a uniform Fabry-P\'erot cavity with region $B$ replaced
  with gain material.}
\label{fig_nonuni_1D}
\end{figure}

\begin{figure}
\centering \includegraphics[width=0.43\textwidth]{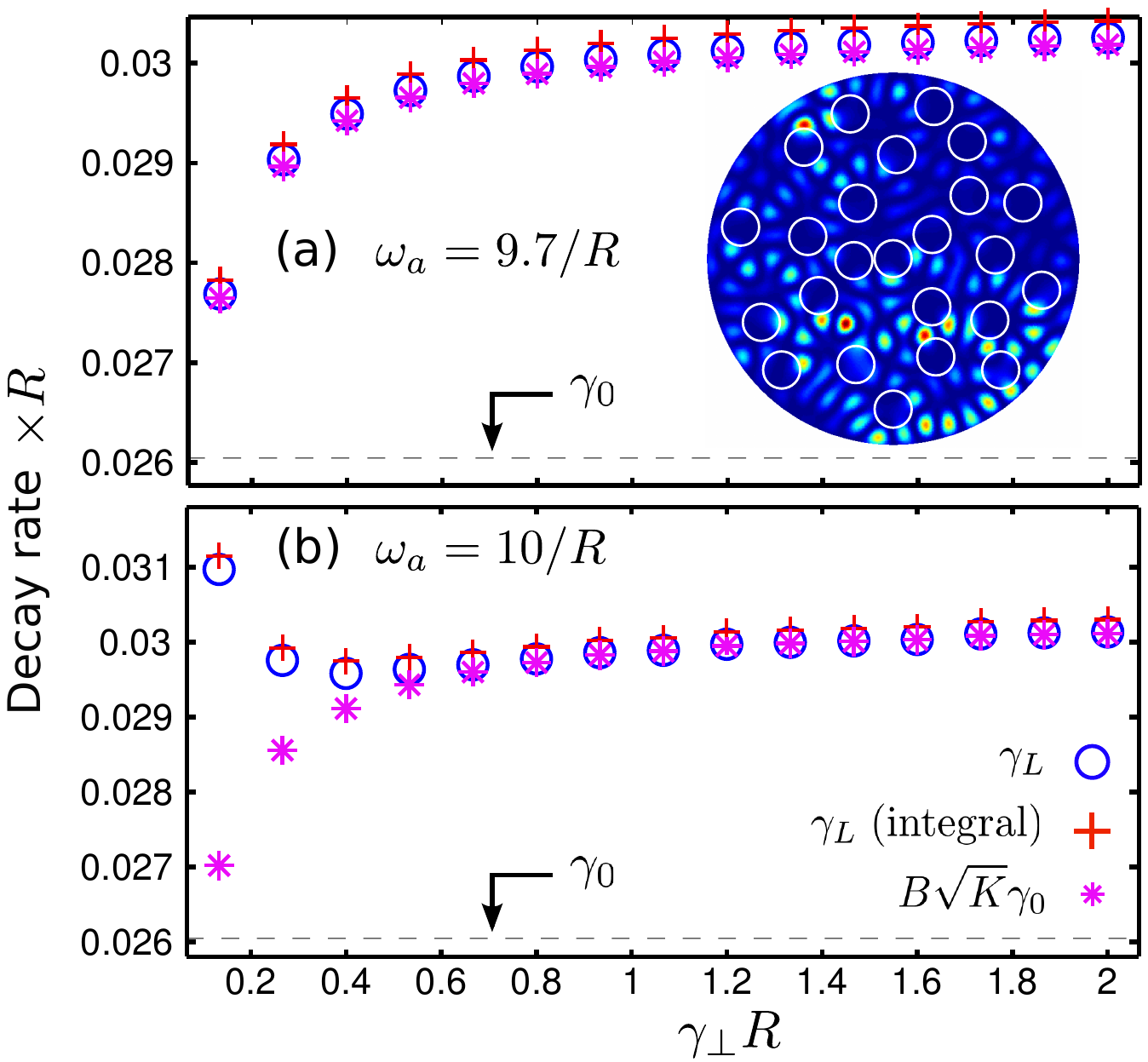}
\caption{(color online) Generalized decay rates versus gain width
  $\gamma_\perp$, for a two-dimensional random laser at threshold.
  The laser cavity consists of a pumped dielectric disk of radius $R =
  1$, background dielectric $n_0^2 = 10$, with 24 randomly-placed air
  holes of radius $0.1$.  The wave equation is solved by the
  finite-element method, and the generalized decay rate $\gamma_L$ is
  computed from the $S$-matrix (blue circles) and from the integral
  approximation Eq.~\ref{gammal_integral complex freq} (red crosses).
  The result of the traditional formula $B \sqrt{K} \gamma_c$ is
  shown for comparison (magenta asterisks), where $K$ is computed from
  the Schomerus formula of Eq.~\ref{Schomerus}. Two values of the gain
  center are used: (a) $\omega_a = 9.7$, for which the lasing mode has
  negligible line-pulling, and (b) $\omega_a = 10$.  Inset: Computed
  mode intensity for the threshold lasing mode.}
\label{fig_twod}
\end{figure}

Fig.~\ref{fig_nonuni_1D} shows the variation of $\gamma_L$ with
$\gamma_\perp$, at threshold, for a non-uniform 1D laser with
spatially inhomogenous dielectric function and pumping.  Over the
entire computed range, the integral formula
(\ref{gammal_integral complex freq}) is again in excellent agreement
with the exact $\gamma_L$ computed from the $S$-matrix.
To compare our $S$-matrix or SALT integral results
for more general cavities to the most complete version of the
``traditional'' results, we combine the Schomerus formula,
Eq.~(\ref{Schomerus}), with an {\it ad hoc} bad-cavity factor. If we
do this, good agreement is observed in Fig.~\ref{fig_nonuni_1D}(a),
when the gain center $\omega_a$ is aligned with one of the passive
cavity resonances and line-pulling is negligible.  In
Fig.~\ref{fig_nonuni_1D}(b), a different choice of $\omega_a$
introduces line-pulling, and the $S$-matrix theory gives significantly
different results, particular for small values of $\gamma_\perp$.

Fig.~\ref{fig_twod} shows an analogous two-dimensional calculation,
for a random laser cavity.  Again, good agreement is observed between
the exact $\gamma_L$ and the integral formula.  The $S$-matrix theory
approaches the traditional result for large values of $\gamma_\perp$
and negligible-line-pulling, and by more than $10\%$ for small values
of $\gamma_\perp$.  This is consistent with the results found in 1D.

\begin{figure}
\centering \includegraphics[width=0.4\textwidth]{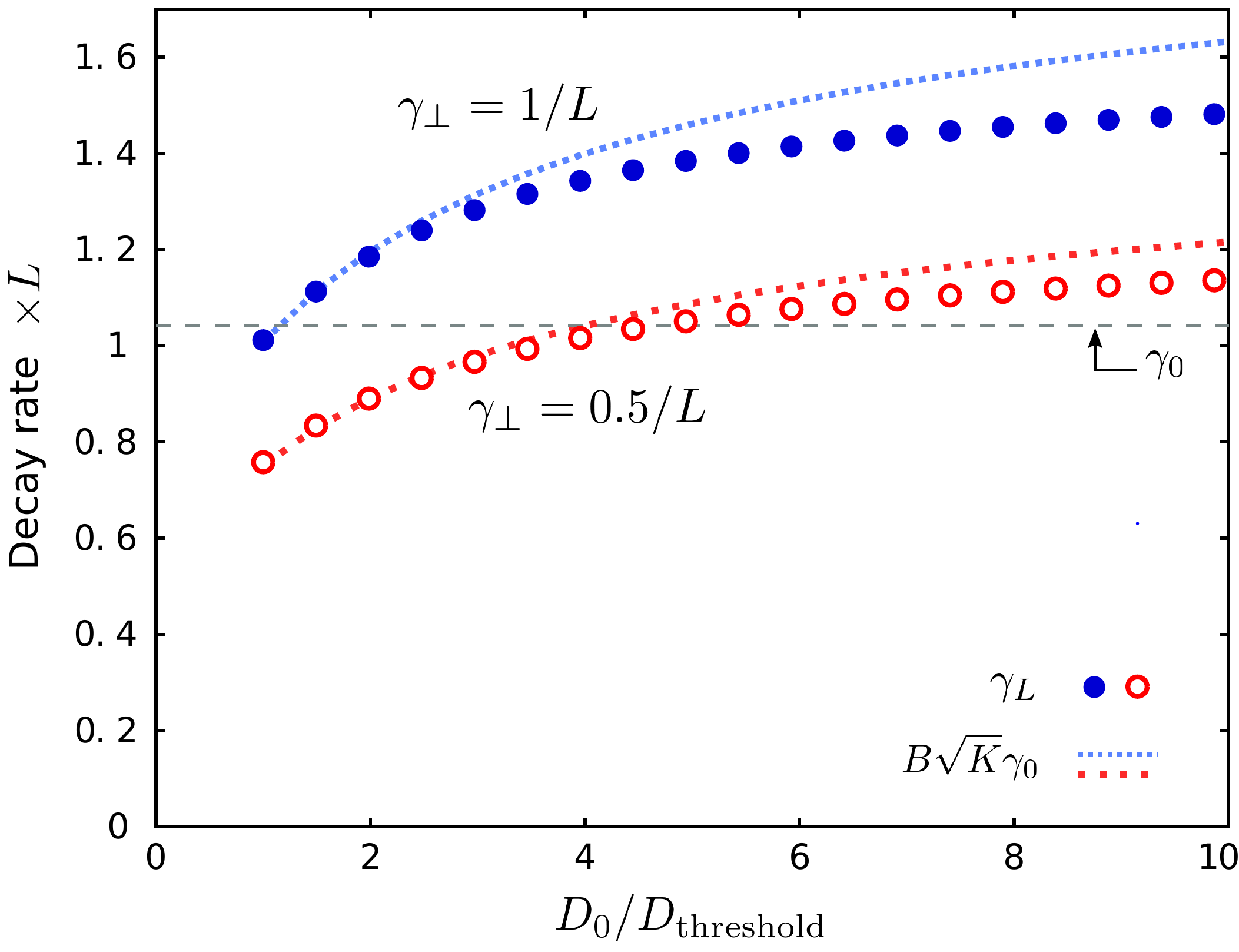}
\caption{(color online) Generalized decay rates for above-threshold
  lasers.  The laser cavity is a slab of length $L$ with a perfect
  reflector on one side.  The gain medium is uniformly pumped, with
  parameters $n_0^2 = 2.3$ and $\omega_a = 100.3/L$ as described via
  Eqs.~(\ref{maxwell bloch eps}) and (\ref{hole burning}).  The pump
  $D_0$ is varied from the threshold value up to $10\times$ threshold.
  The values of $\gamma_L$ are shown for gain widths of $\gamma_\perp
  = 1/L$ (filled circles) and $\gamma_\perp = 0.5/L$ (open circles).
  The dotted lines show the corresponding values of $B \sqrt{K}
  \gamma_c$, where $B$ is the bad-cavity factor and $K$ is the
  Petermann factor of Eq.~(\ref{Schomerus}).  For these parameters,
  the laser is single-mode.}
\label{fig_nonlinear}
\end{figure}

As described in Ref.~\onlinecite{linewidth_prl}, $\gamma_L$ can be
computed above the lasing threshold by using the nonlinear $S$-matrix.
Above threshold, $\epsilon(r)$ is modified by spatial hole burning;
instead of being an independent parameter, the inversion $D$ in
Eq.~(\ref{maxwell bloch eps}) becomes \cite{salt1,salt2,salt3,spasalt}
\begin{align}
  \begin{aligned}
  D(r) &= D_0\; F(r) \left[1 + \sum_\nu \Gamma_\nu|\Psi(r)|^2\right]^{-1}, \\
  \Gamma_\nu &\equiv \frac{\gamma_\perp^2}{\gamma_\perp^2 + (k_\nu-k_a)^2},
  \label{hole burning}
  \end{aligned}
\end{align}
where $D_0$ is the pump strength, $F(r)$ is the spatial profile of the
pump (which is zero in unpumped regions), and $k_\nu$ and
$\Psi_\nu(r)$ are the self-consistently determined frequency and field
function of the $\nu$-th lasing mode (the possibility of multi-mode
lasing is thus explicitly included).  The resulting complex
$\epsilon(r)$ enters into the linewidth theory in exactly the same way
as at threshold: we can obtain the $S$-matrix and hence $\gamma_L$, or
obtain $\Psi_\nu(r)$ and use it directly in the integral formula
(\ref{gammal_integral complex freq}).

Fig.~\ref{fig_nonlinear} shows $\gamma_L$ as a function of $D_0$ for a
1D Fabry-P\'erot cavity.  The value of $\omega_a$ is chosen so that,
at threshold, there is negligible line-pulling; hence the $S$-matrix
and traditional results are in good agreement (as discussed above).
As the pump is increased, the results begin to deviate, up to $4\%$ at
a pump of $10\times$ threshold.  Comparing this to the results of
Ref.~\onlinecite{linewidth_prl}, we conclude that the deviations from
the Schawlow-Townes-Petermann linewidth formula discussed in that
paper was due to the bad-cavity factor.  However,
Fig.~\ref{fig_nonlinear} also shows deviations at large pump
strengths, which cannot be explained by the bad-cavity factor.

In Fig.~\ref{fig_nonlinear}, the variation in the decay rate with
respect to the pump can be intuitively linked to the motion of the
poles and zeros by using Eq.~(\ref{1D ansatz}). Suppose that we have
a single lasing mode at threshold. This corresponds to a pole sitting
on the real frequency axis. As the pump is increased, other poles
and zeros, including the zero associated with the lasing pole,
continue to move up the complex frequency plane. The lasing pole
however remains stationary on the real frequency axis. As a result,
both the prefactor and product terms in Eq.~(\ref{1D ansatz}), and
hence $\gamma_L$, increase with pump, $D_0$. As the pump is
further increased, a second mode turns on and the motions of the
other poles and zeros slow down. This causes the terms in
Eq.~(\ref{1D ansatz}) to remain relatively constant as $D_0$
increases. Thus, $\gamma_L$ increases more slowly with $D_0$.

In conclusion, we have found that the $S$-matrix theory of the laser
linewidth incorporates both the bad-cavity linewidth reduction factor
and the Petermann factor.  For simple cavities, particularly uniform
Fabry-P\'erot cavities with negligible line pulling and close to
threshold, the bad-cavity and Petermann factors can be treated as
independent quantities.  In such systems, we obtain results that are
consistent both with the studies of the bad-cavity factor in
Refs.~\onlinecite{Lax,Haken,Kolobov,Woerdman,Kuppens1,Kuppens2}, and
with Schomerus' generalization of the Petermann factor, without the
bad-cavity factor, in Ref.~\onlinecite{schomerus}.  On the other hand,
in the most general case the bad-cavity and Petermann effects do not
emerge as independent factors, as we saw in
Eq.~(\ref{gammal_integral complex freq}) when expressing the generalized
cavity decay rate in terms of the lasing wavefunction; this was confirmed
in numerical examples with strong line-pulling and spatial hole-burning.
In future work, these deviations will be studied further, with the
goal of developing experimentally feasible laser systems with
anomalous linewidth behaviors.

\section{Acknowledgments}

We would like to thank A.~Cerjan, S.~G.~Johnson, A.~Pick, and
S.~Rotter for helpful discussions.  This research was supported by the
Singapore National Research Foundation under grant No.~NRFF2012-02
and by NSF grant DMR-1307632.

\end{document}